\documentclass[10pt,journal,twocolumn,twoside]{IEEEtran} 

\usepackage{graphicx}
\usepackage{epstopdf}
\usepackage{float}
\usepackage{algorithmic}
\usepackage{array}
\usepackage{amsmath}
\usepackage{amssymb}
\usepackage{mdwmath}
\usepackage{mdwtab}
\usepackage{eqparbox}
\usepackage{stfloats}
\usepackage{fixltx2e}
\usepackage{hyperref}
\usepackage{cleveref}
\usepackage{booktabs}
\usepackage{multirow}
\usepackage{makecell}
\usepackage{subfigure}

\hypersetup{hypertex=true,
            colorlinks=true,
            linkcolor=blue,
            anchorcolor=blue,
            citecolor=blue}
\usepackage{cases} 

\usepackage{upgreek}

\usepackage{xcolor}
\usepackage{makecell}
\usepackage{amsmath}
\usepackage[boxed,ruled,commentsnumbered]{algorithm2e}
\usepackage{url}
\usepackage{cite}
\ifCLASSOPTIONcompsoc
\usepackage[caption=false,font=normalsize,labelfont=sf,textfont=sf]{subfig}
\else

\allowdisplaybreaks[4]

\makeatletter

\renewcommand*{\@opargbegintheorem}[3]{\trivlist
      \item[\hskip \labelsep{\bfseries #1\ #2}] \textbf{(#3):}\ }
\makeatother

\begin{document}
\title
{Less Signals, More Understanding: Channel-Capacity Codebook Design for Digital Task-Oriented Semantic Communication}

\author{Anbang Zhang, Shuaishuai Guo,~\IEEEmembership{Senior Member, IEEE}, Chenyuan Feng,~\IEEEmembership{Member, IEEE}, \\Hongyang Du,~\IEEEmembership{Member, IEEE}, Haojin Li, Chen Sun,~\IEEEmembership{Senior Member, IEEE}, and Haijun Zhang,~\IEEEmembership{Fellow, IEEE} 
\thanks{
Anbang Zhang and Shuaishuai Guo are with School of Control Science and Engineering, Shandong University, China (e-mail: zab\_0613@163.com, shuaishuai\_guo@sdu.edu.cn).

Chenyuan Feng is with the College of Computer Science, University of Exeter, U.K. (email: c.feng@exeter.ac.uk).

Hongyang Du is with University of Hong Kong, China (email: duhy@eee.hku.hk).

Haojin Li and Haijun Zhang are with University of Science and Technology Beijing, China (email: Haojin.li@sony.com, haijunzhang@ieee.org). 

(\emph{*Corresponding author: Haojin Li})

Haojin Li and Chen Sun are with Sony China Research Laboratory, China (email: Haojin.li@sony.com, chen.sun@sony.com).
}
}
\maketitle


\begin{abstract} 
Discrete representation has emerged as a powerful tool in task-oriented semantic communication (ToSC), offering compact, interpretable, and efficient representations well-suited for low-power edge intelligence scenarios. Its inherent digital nature aligns seamlessly with hardware-friendly deployment and robust storage/transmission protocols. However, despite its strengths, current ToSC frameworks often decouple semantic-aware discrete mapping from the underlying channel characteristics and task demands. This mismatch leads to suboptimal communication performance, degraded task utility, and limited generalization under variable wireless conditions. Moreover, conventional designs frequently overlook channel-awareness in codebook construction, restricting the effectiveness of semantic symbol selection under constrained resources.
To address these limitations, this paper proposes a channel-aware discrete semantic coding framework tailored for low-power edge networks. Leveraging a Wasserstein-regularized objective, our approach aligns discrete code activations with optimal input distributions, thereby improving semantic fidelity, robustness, and task accuracy. Extensive experiments on the inference tasks across diverse signal-to-noise ratio (SNR) regimes show that our method achieves notable gains in accuracy and communication efficiency. This work provides new insights into integrating discrete semantics and channel optimization, paving the way for the widespread adoption of semantic communication in future digital infrastructures.
\end{abstract}


\section{Introduction} 
People tend to underestimate what can be accomplished in ten years, but overestimate what can be achieved in two years. Although 5th Generation Mobile Communication (5G) networks are still in the full deployment stage, the research focus in academia and industry has gradually shifted to the exploration of Beyond 5G (B5G) and Sixth Generation Mobile (6G) technologies \cite{1} to address the increasing demand for ubiquitous intelligent communications around 2030. 
This process marks a fundamental shift in wireless communication systems from “connected devices” to “connected intelligence” \cite{2}. 
As we can foresee in the coming decades, the information redundancy and huge resource overhead in modern communication systems motivate us to rethink and reshape the way that we look at communication systems from a semantic-aware perspective \cite{3}. This shift not only reflects the adaptive claims of next-generation (NextG) communication to complex and heterogeneous intelligent scenarios, but also predicts a paradigm evolution: the realization of truly efficient, intelligent, and cost-effective task information interactions through the integration of communication strategies \cite{4} with semantic understanding and inference abilities.

\begin{figure*}[]
\centering
\includegraphics[width=1\linewidth]{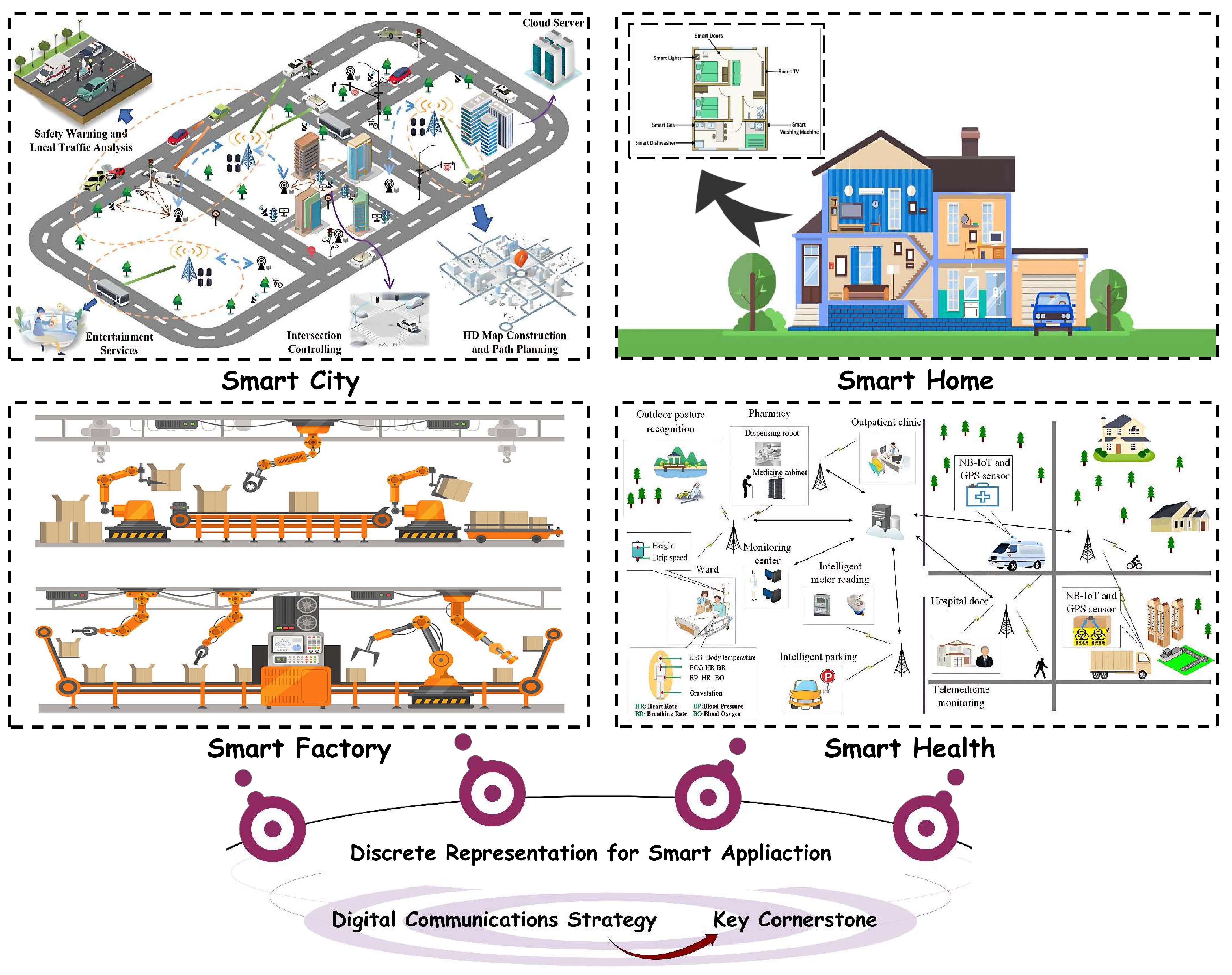}
\caption{Empowering intelligent edge applications with discrete semantic representation.}
\label{fig1}
\end{figure*}

Compared to traditional communication systems that aim to transmit signals accurately, semantic communication, one of the NextG cornerstone technologies, focuses on high-fidelity semantic-level transmission. 
This also shows that NextG systems always have a clear task or goal for data transmission \cite{5}, and has prompted the consideration of task-related semantic awareness at the semantic level.
Therefore, semantic communication is also considered as a task- or goal-oriented communication scheme. This ToSC scheme \cite{7,8} focuses on how the transmitted content accurately conveys the task meaning to the receiver and how the received content effectively influences the decision-making in the desired way. This shift in focus significantly reduces the transmitted redundant information while enhancing robustness to channel variations, thus greatly improving the edge intelligent interaction efficiency.

However, the research on semantic communication experienced a decades-long stagnation, due to the lack of mathematical models for semantic information theory in engineering. Recently, deep learning (DL) \cite{9} has enabled devices with cognitive capabilities and made it possible to realize semantic communication at the engineering level. On the one hand, by considering communication system design as an end-to-end learning task, isolated functions (e.g., source/channel coding and modulation) are replaced by Deep Neural Networks (DNNs), which allow joint optimization of transmitters and receivers for information compression and noise immunity \cite{10} (e.g., deep learning-enabled joint source-channel coding (DeepJSCC) and modulation supported by deep learning). On the other hand, notable successes in the fields of Natural Language Processing (NLP) and Computer Vision (CV) demonstrate the ability of DL to extract the semantics behind text and images. Specifically, this DeepJSCC architecture brings a number of advantages to ToSC systems in terms of efficient task feature extraction. Although DeepJSCC injects the DL advantages into ToSC, its underlying layer still relies on the representation and transmission of continuous semantic features, which poses a critical issue in real-world deployments: most of the current mainstream communication systems use discrete modulation and coding, and the continuous representation generated by DeepJSCC is difficult to be directly compatible with this systems. In low-power, low-complexity edge devices, this mismatch significantly increases a significant increase in hardware implementation complexity and system energy consumption, severely limiting the practicality and scalability of ToSC system. Meanwhile, continuous features usually lack logical structure, which easily leads to semantic distortion under channel interference and affects the task execution effect. Therefore, in the process of advancing the ToSC system toward practical applications, it is necessary to reexamine the interface compatibility between it and the digital communication system, and explore more structured, robust, and transmission-efficient discrete semantic representations to satisfy the intelligent communication deployment requirements in a wide range of scenarios.

Following this trend, ToSC has made breakthroughs in research supporting the efficient connection of massive terminal devices in recent years. For example, with the continuous emergence of ubiquitous intelligent applications, ToSC penetrates into the theoretical and systematic research on typical scenarios such as smart cities, smart homes, smart factories, and smart healthcare, as shown in Fig.\ref{fig1}. These scenarios are usually accompanied by real-time sensing, low-latency decision-making, and task-related information extraction requirements, and the deployment environment is mostly the resource-constrained edge side. The current digital communication technologies have been widely deployed globally and become the closed-loop communication basis for perception-decision-execution in current intelligent systems. However, in the face of task-oriented ubiquitous intelligence needs, the design concept and structure of the traditional bit-level digital transmission paradigm are still outdated. Specifically, the transmission efficiency in existing communication systems is close to the physical limit, but a large amount of information redundancy cannot be identified and compressed, resulting in ineffective channel resource utilization; moreover, the communication system lacks the ability to perceive the task goal, and is unable to proactively filter or optimize the semantic expression, which causes the system to process delays, energy consumption. Therefore, the construction on discrete semantic communication mechanisms based on digital communication foundation and integration of task-driven and channel-aware ability has become a key breakthrough direction for breaking through the communication bottleneck in intelligent scenarios.

Motivated by the discussed above challenges, we systematically analyze the current challenges facing ToSC systems, with a particular focus on the limitations of existing discrete representation strategies. We find that channel-aware symbol discretization and fixed codebook design are the main bottlenecks in achieving robust and efficient ToSC performance. To overcome these issues, we propose a novel channel-aware discrete semantic coding framework based on DeepJSCC (DeepJSCC-CDSC), specifically tailored for low-power edge networks. Our approach introduces the Wasserstein regularization objective to better align learned discrete code activations with the input data semantics and variable channel distribution. This coupling improves robustness to channel variations as well as task-specific accuracy. Experiments under different signal-to-noise ratio conditions show that our method outperforms existing baseline methods in terms of both communication efficiency and semantic task performance. This work provides new insights into the integration of semantic representation learning with wireless channel adaptation, paving the way for scalable, adaptive, and task-aware semantic communication at the network edge.

\section{Challenges and Opportunities in Digital ToSC System} 
\subsection{Rethinking End-to-End Semantics: Why DeepJSCC Falls Short?}

A key issue in ToSC system is how to efficiently extract task-relevant semantic features while minimising task-irrelevant data redundancy. DeepJSCC, as a key architecture for ToSC models, exhibits great potential in addressing technical challenges. Specifically, ToSC system supported by DeepJSCC has a significant advantage in terms of efficient task feature extraction. This framework relies on a joint end-to-end (E2E) approach for pre-training \cite{6}, while leveraging DNNs to extract and transmit task-relevant semantic features. This success is attributed to the learning and fitting capabilities of various neural networks, which break the constraints of mathematical models exchanging semantic information. Although DeepJSCC has achieved significant empirical success in enhancing the E2E performance of semantic communication systems, the following inherent limitations still exist.

With the global deployment of 5G system, research on modern mobile systems typically relies on digital modulation schemes \cite{8}. Moreover, real-world communication infrastructure is designed and guided by Shannon's theorem. Therefore, a ToSC architecture based on DeepJSCC is incompatible with existing standard digital communication systems. Specifically, its encoder output consists of continuous channel input symbols rather than being discretized into constellation symbols. This idealized assumption about transmission requires the analogue modulation or full-resolution constellation modulation, which imposes a significant burden on resource-constrained transmitters and presents implementation challenges for current radio frequency (RF) systems. Additionally, the lack of a structured symbol representation in continuous semantic features results in insufficient robustness when faced with multi-path fading and channel interference in real-world environments. This leads to a significant decline in communication reliability and task accuracy, further limiting the practical application value of DeepJSCC in dynamic real-world environments. Therefore, a critical solution is to explore digital modulation schemes in ToSC system that are more compatible with existing real-world systems.

\subsection{Embracing Discrete Semantics: A Digital-First Communication Paradigm}

Note that the fundamental idea in digital modulation is based on a function that converts real values into discrete constellation symbols. Since this function is non-differentiable, it is impossible to run a gradient descent algorithm to obtain the optimal mapping. To address this challenge, related research has explored quantization-based modulation methods, focusing on optimizing the activation probability in constellation mapping to enable transceivers to process signals more efficiently. In \cite{9252948}, the authors use uniform quantization to uniformly map the DNN outputs into discrete bit sequences, which is then modulated into constellation symbols, addressing the issue of limited constellation size in capacity-constrained IoT devices. In \cite{9791409}, the authors employed non-uniform quantization, providing an example of Binary Phase Shift Keying (BPSK) modulation. An additional DNN with a differentiable Sigmoid function (rather than the non-differentiable step function used in classical BPSK modulation) was deployed to generate the likelihood of constellation symbols, rather than a hard mapping.

However, the above-mentioned modulation methods based on quantizers primarily consider a separated architecture where the modulation module is independent of the coding module. This means that the system cannot adapt to various wireless environments, and transmission performance may be significantly degraded by channel noise. Under the guidance of the DeepJSCC concept, an integrated coding-modulation scheme for digital ToSC systems has been incorporated into the research process. Furthermore, to overcome the limitations imposed by continuous semantic features, recent research has adopted a discretization approach to establish a digital ToSC system supported by vector quantization (VQ) \cite{10101778}, explicitly mapping semantic features to discrete symbols. This naturally aligns with existing digital modulation methods (e.g., BPSK, QPSK, QAM, etc.), to achieve a more compatible digital communication system and more expressive discrete semantic encoding.

In addition, structured discrete representations improve robustness against noise and channel impairments and facilitate end-to-end optimization using advanced generative models like vector-quantization variational autoencoders (VQ-VAE). For example, in \cite{8}, the authors have proposed a ToSC scheme that discretizes feature representations before digital modulation, achieving better task inference accuracy and lower latency. Furthermore, a mask-based VQ-VAE \cite{10101778} is introduced that enhances noise robustness by encoding semantic features using a shared discrete codebook. In \cite{10110357}, a $\beta$-variational autoencoder ($\beta$-VAE)-based semantic communication system is developed to enable interpretable feature selection and improved robustness to semantic noise. These efforts mark a fundamental shift in communication paradigms—from "symbol transmission" to "semantic knowledge exchange."

\begin{figure*}[]
\centering
\includegraphics[width=1\linewidth]{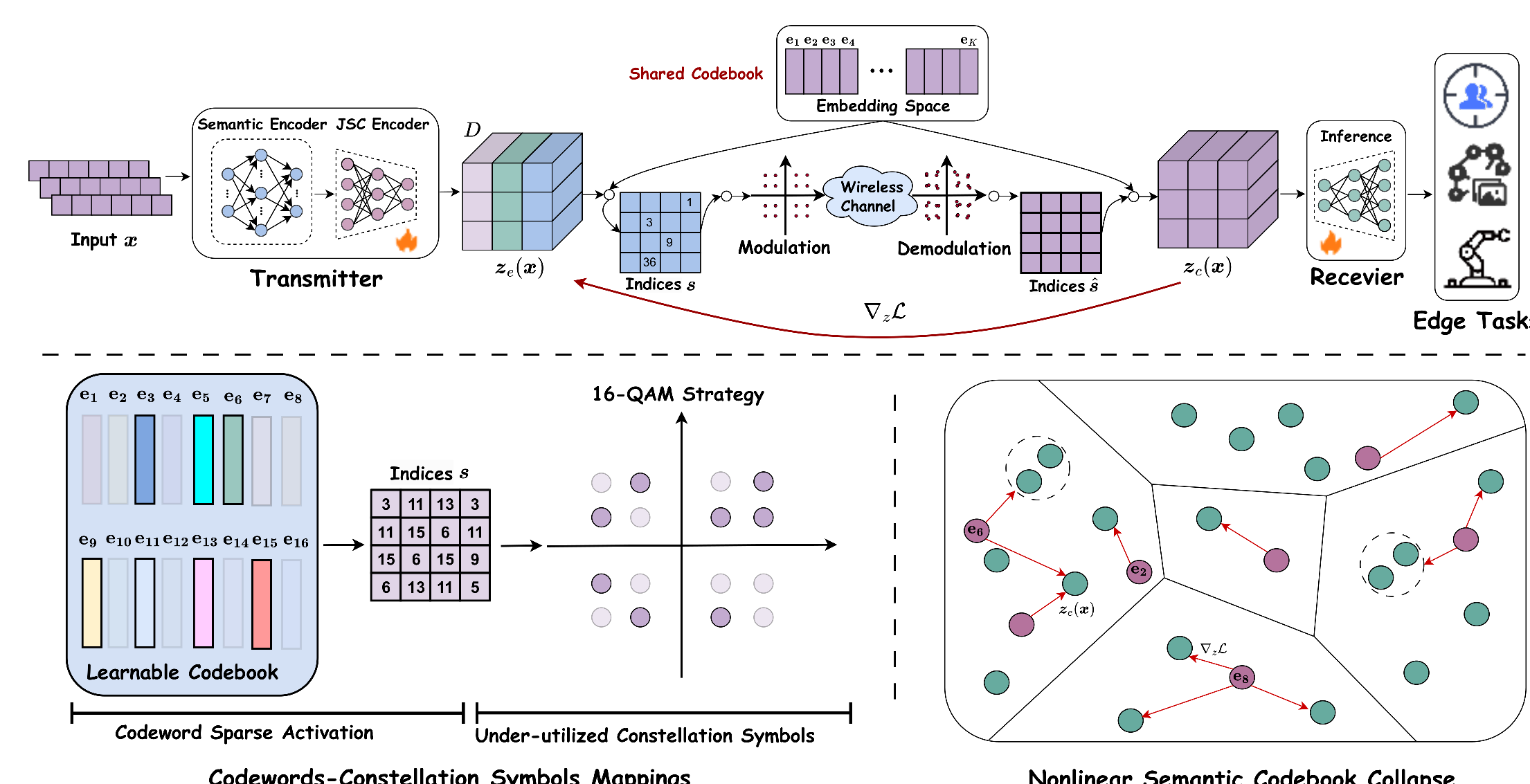}
\caption{Sparse codebook activation and under-utilized constellation symbols: an illustration of channel-aware discrete semantic coding.}
\label{fig2}
\end{figure*}

\subsection{From Promise to Practice: Navigating the Barriers to Discrete Semantic Communication}
Despite the great potential shown by digital ToSC system, its practical deployment still faces several key challenges, as shown in Fig. \ref{fig2}: 
\begin{itemize}
\item \emph{Guaranteeing the stability on nonlinear semantic mappings:} Nonlinear mappings from semantic space to discrete symbol space present significant stability challenges during training, especially when complex generative models such as vector quantization models are involved. Utilizing stable optimization methods and loss functions can ensure efficient and reliable training.
\item \emph{Joint task-channel optimization strategies:} channel conditions in real communication systems change dynamically, and achieving an optimal balance between task accuracy and channel robustness has become a key research goal. In the process of exploring end-to-end optimization strategies, fully considering the awareness ability to channel conditions can further enhance the adaptivity and generalization ability in different channel scenarios.
\item \emph{Codebook Collapse Issues:} VQ-support models often suffer from codebook collapse, i.e., a large number of semantic features are mapped to a few codewords. This imbalance significantly reduces the communication capacity and semantic expressiveness. The development in novel training strategies is crucial to alleviate this problem and achieve effective Digital ToSC systems.
\end{itemize}

Although digital ToSC has obvious advantages over continuous DeepJSCC methods, it is still crucial to address the above technical frontier challenges. Future research needs to continue to promote theoretical and practical innovation in order to drive semantic communication from theoretical exploration to reliable practical deployment.

\section{Discrete Mapping Scheme: Bridging Semantic Precision and Channel Reality} 
\begin{figure*}[]
\centering
\includegraphics[width=1\linewidth]{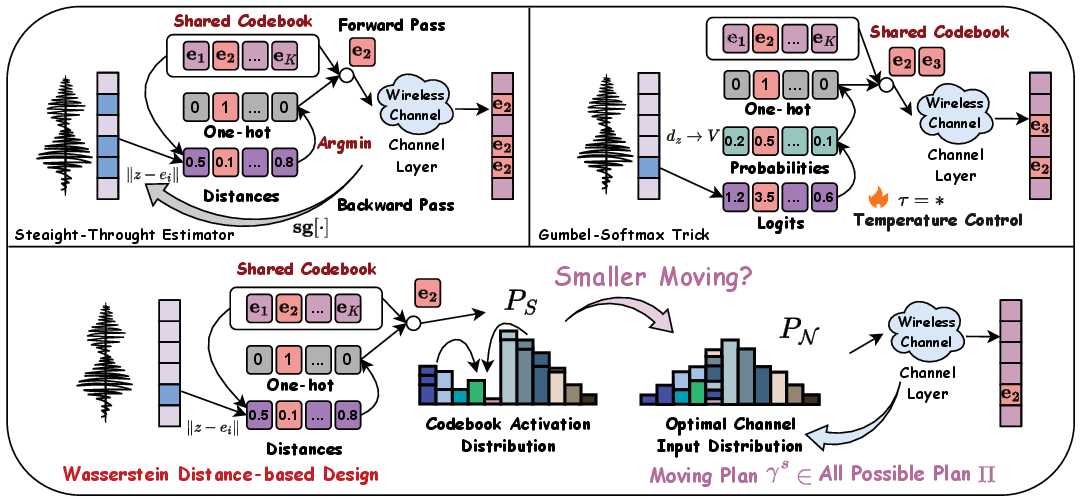}
\caption{From hard choices to smooth alignments: STE, Gumbel-Softmax, and Wasserstein distance for discrete semantic designs.}
\label{fig3}
\end{figure*}

To achieve efficient, robust digital ToSC in low-power edge intelligence scenarios, as shown in Fig. \ref{fig3}, we design a discrete mapping scheme that ensures semantic accuracy while adapting to complex channel conditions has become a key research issue. And we analyze two typical discrete mapping schemes currently in usage, identify their limitations.

\subsection{STE-driven Vector Quantization}
In related researches on ToSC systems, several methods based on VQ combined with Straight-Through Estimator (STE) technology are primarily used to address training challenges arising from non-differentiable operations during the discretization process. 
To address this issue, the STE method performs hard selection during forward propagation to ensure the effectiveness of discretizations, while approximating this operation as an identity mapping during backward propagation, directly propagating gradients back to the continuous feature generation module, thereby ensuring the smoothness and effectiveness of the overall training process. 

However, while STE-based vector quantization schemes effectively address the trainability issues of discretization, core optimization objectives on ToSC system still focus on preserving the expressive power and sparsity of discrete representations, neglecting the dynamic changes in channel conditions. In real-world wireless communication environments, channel conditions are often unpredictable and constantly changing, such as fluctuations in signal-to-noise ratio, and multi-path interference, all of which significantly impact transmission quality and final task inference performance. The lack of explicit modeling of channel characteristics means that this scheme cannot promptly adjust codeword activation probabilities and feature distributions when faced with dynamic or adverse channel conditions, ultimately leading to insufficient transmission robustness, reduced task inference accuracy, and even the need for additional redundant designs to maintain performance, further increasing system complexity and energy consumption. While this channel-unaware design approach simplifies system training and implementation, it has increasingly revealed obvious limitations in real-world low-power edge intelligence applications, particularly in scenarios requiring high reliability and adaptability.

\subsection{Gumbel-Softmax Empowered Quantization}
Another typical traditional discrete representation optimization strategy, namely the vector quantization method based on Gumbel-Softmax, has been proposed to achieve end-to-end trainable optimization of discrete features.
Unlike conventional hard one-hot sampling, Gumbel-Softmax introduces random noise into the logits and employs a temperature parameter to control the degree of smoothness. This mechanism allows the representation to smoothly approximate discrete selections during training, preserving the required sparsity for discrete semantics while maintaining gradient flow throughout the entire optimization process.

The primary advantage of this method lies in its ability to significantly enhance training stability and improve inference accuracy, while simultaneously ensuring that the semantic content is effectively maintained in the transmitted representation. The soft sampling strategy also mitigates the risk of gradient vanishing or explosion that often arises in non-differentiable discrete selection scenarios, making it more suitable for deep neural network-based joint optimization frameworks. However, despite these advantages, this scheme inherently focuses on optimizing the discrete feature representation itself, without considering dynamic channel conditions or variations that frequently occur in real-world wireless communication scenarios. The robustness of the model relies mainly on introducing additional encoding redundancy, rather than adapting the discrete codeword activation probabilities based on real-time channel states such as fluctuating signal-to-noise ratios (SNR), fading effects, and multi-path interference. This “channel-agnostic” design approach, while simplifying the optimization objective, severely limits the flexibility and robustness of the scheme in practical deployments. This underscores the urgent need for future research to explore channel-aware discrete representation learning strategies.


\subsection{Channel-Aware Discrete Semantics Design}
Currently, mainstream discrete mapping optimization schemes generally adopt a channel-postprocessing or channel-unaware design approach. These methods typically optimize the representation and coding strategy of discrete features under idealized channel conditions, and then address channel disturbances through the addition of redundancy or post-processing techniques. While this strategy may perform well under conventional channel conditions, it often struggles to maintain high stability and accuracy in real-world wireless environments due to its inability to adjust encoding schemes in real-time based on channel states, significantly limiting its practical application value.

To address this core challenge, this paper proposes a novel discrete mapping scheme centered on channel awareness. Unlike traditional methods, this scheme explicitly incorporates channel statistical characteristics into the optimization process of discrete feature probability distributions for the first time. By introducing the Wasserstein distance as a regularization constraint, the model dynamically aligns the distribution of discrete codewords with the optimal channel input distribution, thereby achieving deep synergistic optimization between feature representation and channel state. Specifically, the Wasserstein regularization mechanism guides the encoder during training to minimize the gap between the current discrete codeword activation distribution and the ideal distribution designed for the channel conditions. This design enables the encoder to autonomously adjust the activation probabilities of codewords based on real-time channel conditions, achieving rapid adaptation to complex wireless environments. This adaptive capability effectively mitigates the impact of channel fluctuations on semantic transmission, reduces reliance on additional redundancy or conservative designs, and further optimizes transmission performance.

Additionally, the channel-aware discrete mapping strategy tightly couples task information, channel conditions, and encoding strategies into a unified framework. This deep integration fundamentally breaks away from the traditional channel-postprocessing approach, directly incorporating channel information into the learning process of discrete representation. Especially in low-power, dynamically changing edge intelligence application scenarios, this solution can sustainably and stably support high-performance task inference under resource constraints, demonstrating significant application potential.

\section{Experiments and Discussions}
In this section, we compare our proposed scheme with other DL-based baseline scheme. Image classification task is chosen as an example to demonstrate the effectiveness of the proposed scheme, which is to develop an end-to-end learning framework to extract channel-aware latent representations for transmission. It is noteworthy that the proposed approach is not specified for the image classification task.

\subsection{Experimental Settings}
\emph{Compared Methods:} We evaluate our proposed method against three state-of-the-art ToSC baselines, described as follows:
\begin{itemize}
\item \textbf{DeepJSCC-G:} Based on NECST \cite{Choi2018NeuralJS}, this method utilizes a DNN to map input data to channel symbols and jointly learns both encoding and decoding processes.  In our implementation, the encoder outputs are modeled as Gaussian-distributed signals and directly transmitted over an AWGN channel. To distinguish this variant from the original NECST, we refer to it as DeepJSCC-G, and employ the cross-entropy loss for training.
\item \textbf{DeepJSCC-VIB:} This method builds on VFE \cite{7}, a learning-based ToSC framework that leverages the variational information bottleneck (VIB) principle to minimize communication cost. It introduces sparsity-promoting priors and dimension pruning, enabling symbol transmission over reduced dimensions. Our implementation follows the same training strategy and is denoted as DeepJSCC-VIB for clarity.   
\item \textbf{DeepJSCC-RIB:} Derived from DT-JSCC \cite{8} scheme, this baseline method incorporates a robust information bottleneck (RIB) to enhance resilience against channel variations, achieving a balance between informativeness and robustness. We adopt the same training scheme and refer to this version as DeepJSCC-RIB. 
\item \textbf{DeepJSCC-CDSC:} Our proposed DeepJSCC-CDSC scheme integrates channel awareness into discrete semantic coding via a Wasserstein-regularized objective. By aligning codeword activation with optimal channel-adaptive distributions, DeepJSCC-CDSC enhances semantic fidelity and robustness, enabling efficient and accurate task inference in dynamic wireless environments.
\end{itemize}

To enable fair and consistent comparisons, we standardize the experimental settings across all evaluated methods, including our proposed Deep-CDSC. In particular, we use an identical discrete codebook structure and employ the same $K$-ary quadrature amplitude modulation ($K$-QAM) scheme to convert discrete latent representations into channel symbols. Additionally, all models utilize the same neural network backbone and are constrained to equivalent computational complexity and memory budgets, effectively simulating realistic deployment scenarios on resource-constrained edge devices.

\subsection{Dataset and Neural Network Architecture}
To evaluate model performance, we employ the CIFAR-10 dataset, which comprises 60,000 color images of size 32×32 pixels, evenly distributed across 10 distinct categories. Specifically, the dataset is split into 50,000 images for training (with 5,000 samples per class) and 10,000 images for testing. During the training phase, we incorporate standard data augmentation strategies such as random cropping and horizontal flipping to improve the model's generalization capability and robustness. Leveraging this dataset, we construct the model backbone using a combination of convolutional layers and ResNet blocks, enabling effective feature extraction and representation learning tailored for digital ToSC systems.

\subsection{Performance Evaluation}
To evaluate the task inference capability, we tested various codebook sizes ($K=16$, $64$, and $256$) corresponding to 16-QAM, 64-QAM, and 256-QAM, along with two codeword dimensions ($D=64$ and $512$). All models were trained at 12 dB and tested under diverse SNR conditions ranging from 4 dB to 20 dB.

\begin{figure*}[!t]
        \centering
        \subfigure[Codebook size $K=16$.]{{\label{fig4a}}\includegraphics[width=0.325\linewidth]{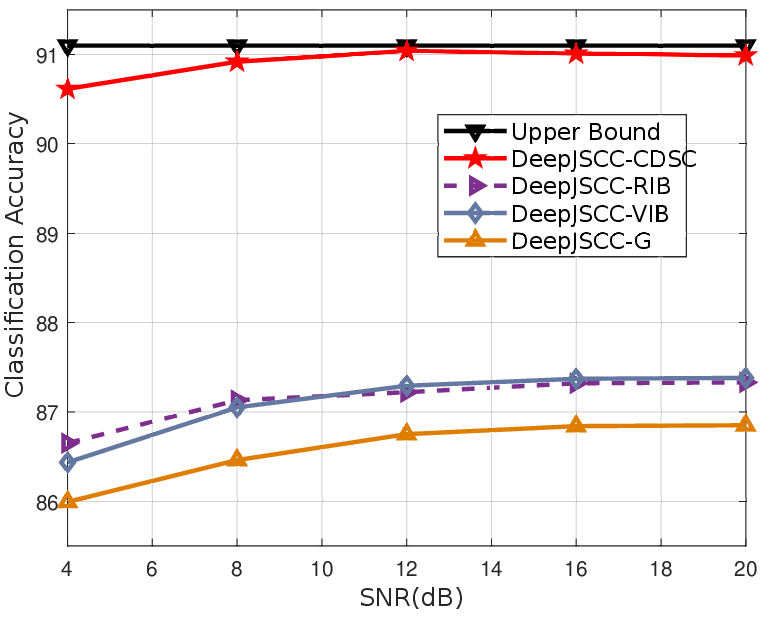}}
        \subfigure[Codebook size $K=64$.]{{\label{fig4b}}\includegraphics[width=0.32\linewidth]{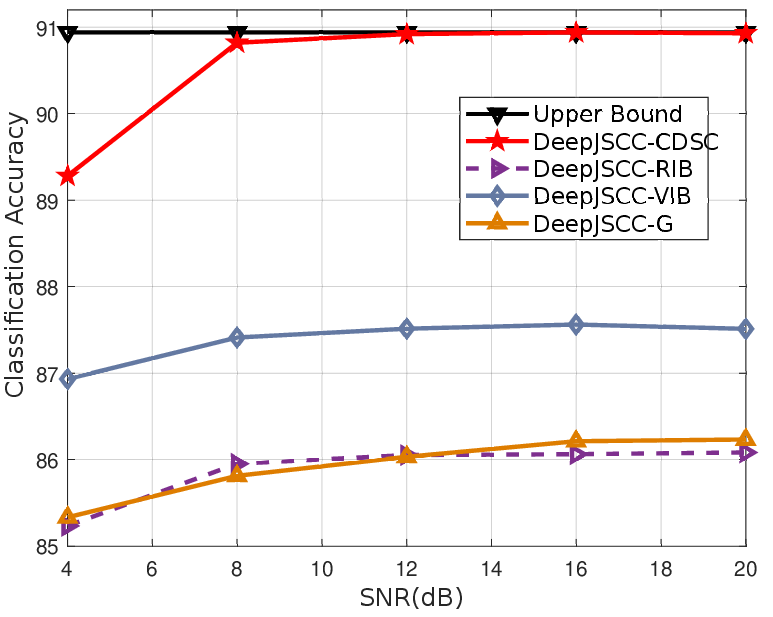}}
        \subfigure[Codebook size $K=256$.]{{\label{fig4c}}\includegraphics[width=0.32\linewidth]{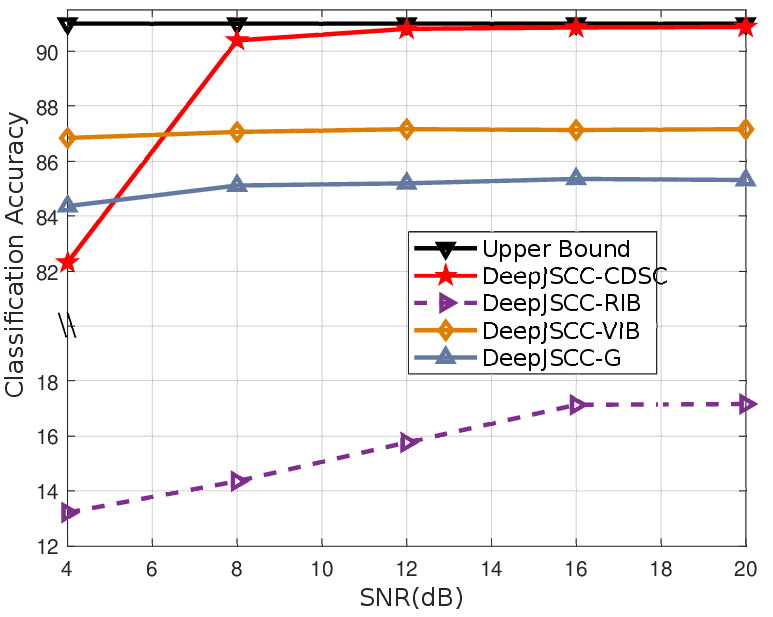}}
        \caption{Task performance comparison versus SNR, where Codebook size $K=16$, $K=64$, $K=256$ and Codeword dimension $D=64$.}
        \label{fig4}
\end{figure*}

\begin{figure*}[!t]
        \centering
        \subfigure[Codebook size $K=16$.]{{\label{fig5a}}\includegraphics[width=0.32\linewidth]{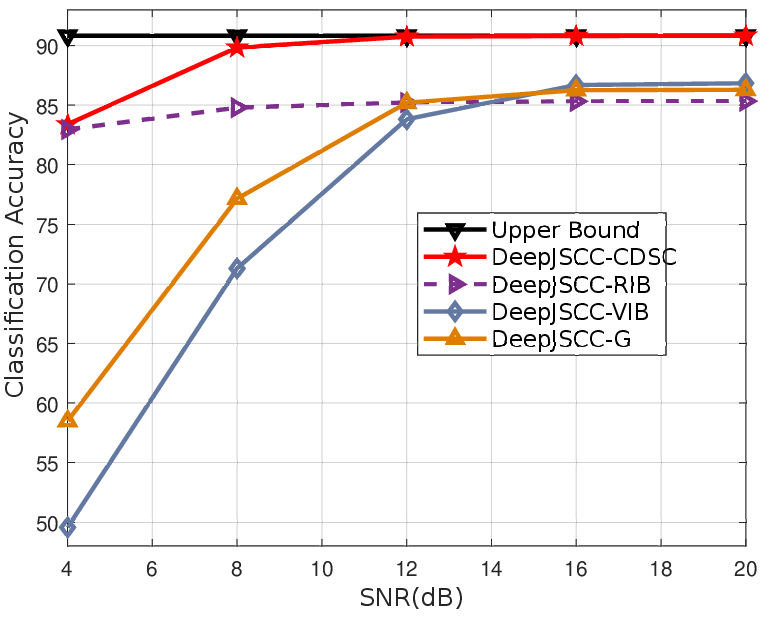}}
        \subfigure[Codebook size $K=64$.]{{\label{fig5b}}\includegraphics[width=0.32\linewidth]{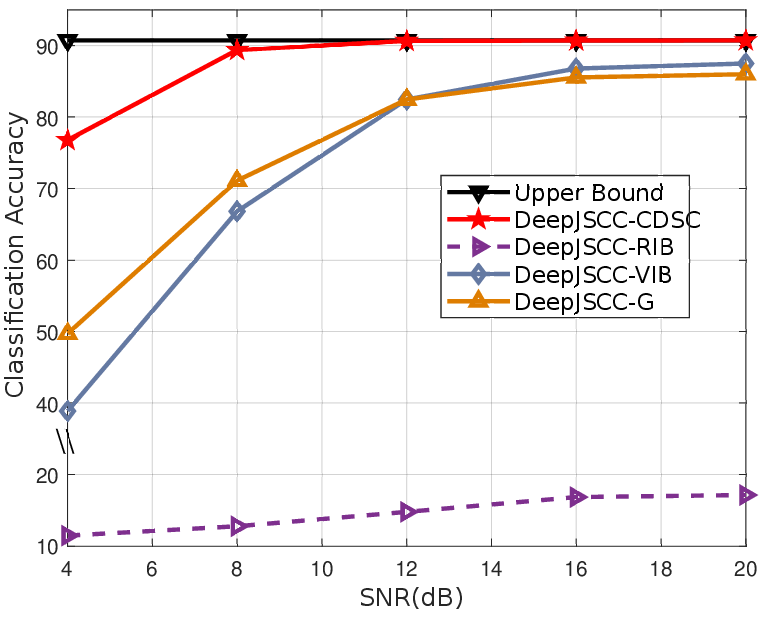}}
        \subfigure[Codebook size $K=256$.]{{\label{fig5c}}\includegraphics[width=0.32\linewidth]{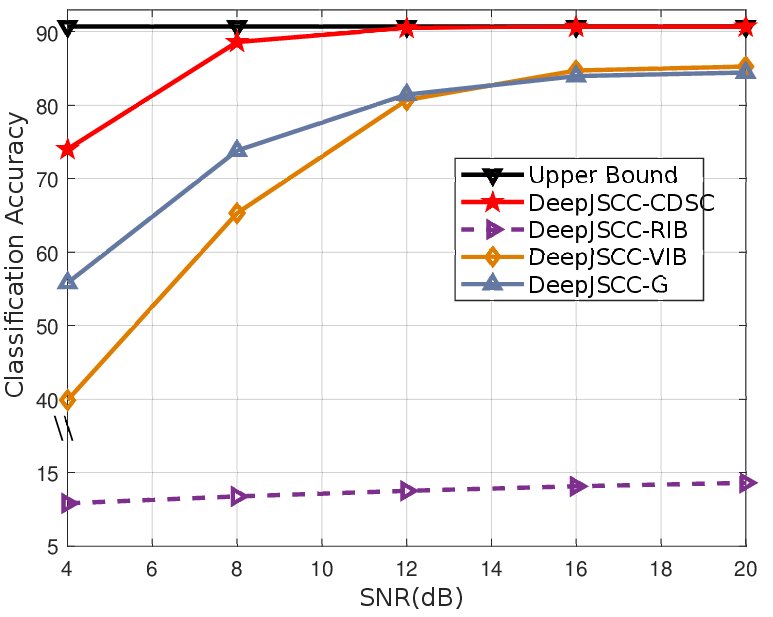}}
        \caption{Task performance comparison versus SNR, where Codebook size $K=16$, $K=64$, $K=256$ and Codeword dimension $D=512$.}
        \label{fig5}
\end{figure*}

\emph{Compact Coding Resilience:}
As shown in Fig. \ref{fig4a} and Fig. \ref{fig5a}, DeepJSCC-CDSC consistently achieves the highest accuracy across all SNR levels when using a compact codebook ($K=16$). Even at low SNR (4 dB), it maintains near-optimal performance with only minor fluctuations. In contrast, baseline methods suffer significant accuracy drops under noisy conditions, revealing their heavy dependence on favorable channels. However, as the codeword dimensionality increases, the performance of all schemes is affected. However, the degradation of DeepJSCC-CDSC is minimal. The other comparison schemes show a significant performance degradation, which indicates that high-dimensional coding is more susceptible to channel noise.  Even at low SNR, DeepJSCC-CDSC can also maximize the important information preservation and shows strong channel robustness.

\emph{Adaptability with Larger Codebooks:}
When the codebook is extended to $K=64$ in Fig. \ref{fig4b} and Fig. \ref{fig5b}, regardless of the dimensions on $D=64$ and $D=512$, DeepJSCC-CDSC can still maintain over 90\% at medium to high signal-to-noise ratios and remain robust even at 4dB. At the same time, the baseline showed significant performance degradation, especially under high-dimensional conditions. The codebook extension enhances the expressive power, but the redundant representation at $D=512$ exacerbates more severe channel interference. Especially when the channel conditions are poor, some schemes (such as DeepJSCC-RIB and DeepJSCC-G) experience a rapid decline in accuracy due to a lack of sufficient channel awareness. In contrast, the proposed DeepJSCC-CDSC scheme maintains high classification accuracy by matching channel capacity with a compact coding strategy to minimize redundant information interference. For example, DeepJSCC-CDSC reduced the negative impact of increasing codebook freedom degrees ($D=512$) by 63.46\% (compared to DeepJSCC RIB), 36.03\% (compared to DeepJSCC-VIB) , and 25.2\% (compared to DeepJSCC-VIB).

\emph{Advantages in Extreme Configurations:}
With a large codebook ($K=256$) in Fig. \ref{fig4c} and Fig. \ref{fig5c}, DeepJSCC-CDSC outperforms best, achieving stable accuracy above 90\% regardless of codeword dimension. In contrast, all baselines struggle to handle sparse codeword activations and increased transmission complexity, leading to dramatic accuracy loss and severe instability, particularly for DeepJSCC-RIB. This confirms that  the large codebook size brings stronger information expression capability to the model, but also amplifies the codebook sparse activation problem. In particular, under the worse channel conditions, the increased freedom may not bring better performance, but exacerbate the channel interference and overfitting problems.


\section{Future Research Directions}
Although our proposed channel-aware discrete semantic coding framework has demonstrated strong performance in low-power edge intelligence scenarios, there remain several important research directions worthy of further exploration:
\subsection{Joint Optimization with Embodied Intelligence Systems}
Future work can explore integrating channel-aware discrete ToSC with embodied intelligence systems to achieve holistic co-optimization across perception, semantic understanding, control, and feedback. Embodied intelligence emphasizes that agents continuously interact with the physical environment to build dynamic, adaptive cognitive models. This close coupling requires communication systems to go beyond static or single-task optimization strategies. By jointly designing semantic coding with perception and action policies, the system can dynamically adjust discrete encoding strategies in real time based on channel states, task priorities, and environmental changes. Such integration enables not only efficient and reliable information transmission but also improves task-driven decision-making and environmental adaptability. This innovative direction can support complex applications such as autonomous robots, smart manufacturing, remote surgery, and collaborative drone fleets, pushing semantic communication from mere transmission optimization toward task-driven intelligent perception and action in a multi-dimensional manner.

\subsection{Scalable Multi-Task and Multi-Modal Applications}
While the current framework primarily focuses on single-task inference, extending it to multi-task and multi-modal scenarios is a promising research direction. For instance, autonomous driving requires simultaneous processing of lane detection, pedestrian recognition, and traffic sign understanding; in intelligent factories and collaborative robotics, systems often need to fuse visual, auditory, and tactile signals for complex reasoning and control. Such extensions demand more flexible encoding strategies and dynamic prioritization among different tasks to optimize semantic representation and resource allocation efficiency. Future studies can investigate shared or task-specific discrete codebooks, task-priority scheduling mechanisms, and cross-modal semantic compression techniques to improve system generalization and adaptability across diverse edge applications.

\subsection{Privacy- and Security-Aware Semantic Coding}
Although discrete coding inherently enhances security through sparsity and compactness, the growing sophistication of attack techniques necessitates deeper integration of advanced privacy-preserving mechanisms into semantic coding design. Future research could incorporate methods such as adversarial training, defensive randomization, and differential privacy to strengthen resistance against model inversion, eavesdropping, and inference attacks. Moreover, combining physical-layer security techniques (e.g., artificial noise injection, random precoding) with semantic coding can further boost end-to-end privacy guarantees. Such an integrated approach can achieve a balance among communication efficiency, inference accuracy, and privacy protection, addressing stringent security requirements in sensitive applications such as healthcare, finance, and personalized services.

\section{Conclusions}
This article provides a comprehensive review of the latest developments, challenges, and opportunities in semantic communication. A channel aware discrete semantic communication framework is systematically studied and proposed for digital task oriented semantic communication in low-power edge intelligence scenarios, which explicitly aligns the distribution of semantic codewords with dynamic channel characteristics. The performance evaluation verified the excellent adaptability, compact coding advantages, and ability to support reliable and efficient semantic communication in practical edge intelligence applications of the proposed DeepJSCC-CDSC scheme. In addition, this article systematically introduces the challenges and key issues faced in current research on digital ToSC, and outlines future research directions. We believe that this work has the potential to provide new ideas for continuous innovation in the field of discrete semantic communication, and to contribute to the development and evolution of intelligent wireless networks.

\bibliographystyle{IEEEtran} 
\bibliography{bib}

\end{document}